# Spreadsheet on Cloud – Framework for Learning and Health Management System

[1]K.S. Preeti,[2]Vijit Singh, [3] Sushant Bhatia , [4]Ekansh Preet Singh, [5]Manu Sheel Gupta

WZ-106/73 Rajouri Garden Extension

New Delhi – 110027, India

[1] Independent Developer, Email Add: kspreeti@nsitonline.in

[2,3,4] Software for Education, Entertainment and Training Activities, Email Add: manu@seeta.in,vijit@seeta.in, sushant@seeta.in, ekansh@seeta.in

**ABSTRACT**

Cloud Computing has caused a paradigm shift in the world of computing. Several use case scenarios have been floating around the programming world in relation to this. Applications such as Spreadsheets have the capability to use the Cloud framework to create complex web based applications. In our effort to do the same, we have proposed a Spreadsheet on the cloud as the framework for building new web applications, which will be useful in various scenarios, specifically a School administration system and governance scenarios, such as Health and Administration. This paper is a manifestation of this work, and contains some use cases and architectures which can be used to realize these scenarios in the most efficient manner.

## I. INTRODUCTION

With industry analysts such as Gartner placing cloud computing amongst the top 10 strategic technologies for 2011, it is not a surprise that the world is now looking at this as the next buzz word. [1] In this context, as enterprises and organizations across the world gear up to embrace the technology, it is up to the creators of software to localize Cloud Computing and customize it in a manner suitable to the situation in which it is being deployed. It has also become imperative to import the existing systems into the corresponding cloud versions and utilize them to expand our presence in the cloud.

One such essential platform is that of the Spreadsheet. In the business world, the enterprises and education system, the Spreadsheet has become an indispensable tool of productivity and organization. The Spreadsheet is not just an effective information system, but it is also a strong and reliable framework for building applications. While the movement of Spreadsheet towards the cloud has started taking place in various forms such as Google Docs and Editgrid, what remains to be done is using the Spreadsheet framework over the Cloud to create innovative services which utilize the mathematical and programming capabilities of the activity and at the same time leverage upon the collaborated environment of the cloud. Not only this, these services must also have the capability to be customized for the typical use-cases. Through this paper, we examine various such scenarios.

Our work with One Laptop Per Child (OLPC Inc.) [2] led us to develop a novel Spreadsheet activity, SocialCalc for desktops. As an experiment, in order to bring Computer Supported Collaborative Learning into the education system, we imported the Activity in to the Cloud. Once this was achieved, we examined various scenarios, apart from education where Collaboration over the Spreadsheet framework could be utilized to create fresh innovative software. Our work in this sphere brought us to the Indian context and hence, to the idea of using Collaborative Spreadsheets for developing customizable web applications such as a School system. In the course of the paper, we have examined thisconcept and also explained briefly how the Collaboration over Spreadsheet activity was achieved. Also, we have investigated a new architecture for deploying these applications successfully over the cloud and allow multiple users to collaborate and work.

## II. INTRODUCTION TO SOCIALCALC SPREADSHEET ACTIVIY

One Laptop Per Child (OLPC) is an organization dedicated to create educational opportunities for the world's poorest children by providing each child with a rugged, low-cost, low-power, connected laptop (XO) with content and software designed for collaborative, joyful, self-empowered learning. As community engineers associated with this unique proposition, we have constantly evolved our programming skills to align ourselves with the mission statement and develop software for educational purposes.

SocialCalc is a spreadsheet activity developed for functioning in the Sugar environment [3], OLPC's software paradigm. Initially coded by Dan Bricklin, Founder and President of Software Garden Inc. for Socialtext, Inc. [4], the OLPC part was started by Manusheel Gupta, Managing Director of SEETA with K.S. Preeti, community friend of SEETA and alumnus from NetajiSubhas Institute of Technology and Luke Closs from Socialtext Inc. under the guidance of Walter Bender, Oversight Board Member at Sugar Labs.

The main idea of the Spreadsheet activity is to include features that would enable children to make easy use of the typical features of Spreadsheet activities such as organization, graphing and simple calculations in their respective languages. The main features of this spreadsheet activity are:

- Tabulation
- Organization
- Graphing and Calculation
- Localization in different languages
- Multi-user editing over the mesh network
- Ability to read and edit single sheet Excel 1997-2003 (.xls), Lotus (.wk4) and other popular spreadsheet files
- Optimization in saving of sheet data.
- Collaboration over the Cloud
- Chat integration

Over time, SocialCalc has grown to become an innovative platform over which we have experimented on several accounts of collaborative learning. The basic framework of the application is as follows:

- Application – Spreadsheet activity called SocialCalc, written in JavaScript
- Platform – Python, integrated through XOCOM Library
- Infrastructure – XO Laptop and School server

During the course of this development, we also created a library, XOCOM for running DHTML activities in the Sugar development environment [5]. The basic utility of the package is in its ability to help integrate JavaScript codes with Python codes, hence, ensuring a flexible and robust communication between both. We have utilized this library for all functionalities, including Collaboration.

### III. TAKING SPREADSHEET TO THE CLOUD

In order to take our Spreadsheet Activity to the cloud, we devised an architecture which enables the application to reside on the school server, common to all the XO laptops. All the systems integrated to the cloud access the application on the browser, while the server handles operations such as saving, etc. Though SocialCalc was ready to be used by individual browsers on their XOs, additional infrastructure was needed in order to support collaboration. Changes were introduced in the Python as well as JavaScript parts with XOCOM acting as the base, to create the infrastructure.

To put into effect this use-case, we first accomplished the same on an established Cloud server, that is, the Google App Engine. The Python code was used for the server side scripting, while the JavaScript code running on the Browser acted as the main activity. This application was named SocialCalcNet.

### IV. FEATURES OF SPREADSHEET ON THE CLOUD

A. **Login -** User login is required to keep record of the sheets that a user creates. SocialCalcNet provides the user an option of logging in either using his Googleaccount or creating his own account with the application. Whenever a user logs in, a new session is created which identifies each user.

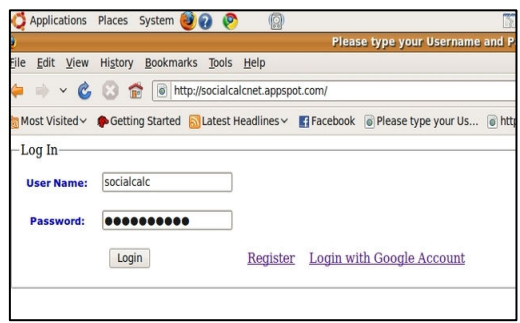

**Fig 1: Login Page**

B. **Account Details -** This displays the account details of the user and also provides the options for further actions – new sheet, load a previous sheet, edit account details, etc.

C. **Edit Account -** Users can edit their accounts created with the application. They can change username or password. However, Google account details cannot be changed.

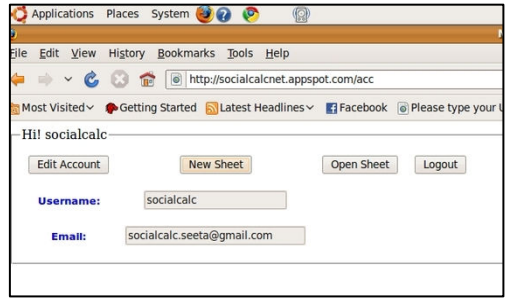

**Fig2: Login Details**

D. **Socialcalc Sheet -** Users can create new sheets and work on them. A tab by the name Options was added to the application. This tab provides options to the user such as saving the current sheet, opening a previous sheet or logging out of the current account.

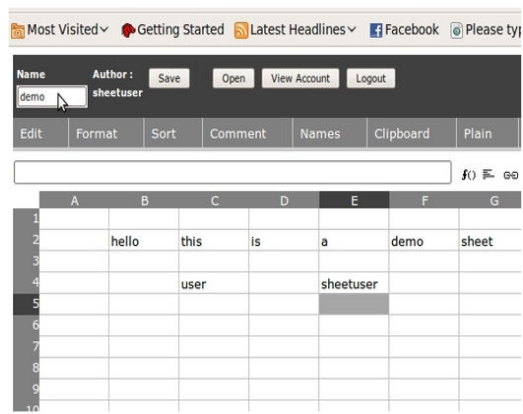

Fig 3: Editing a Sheet

E. **Editing a Sheet -** Users can also open previously saved sheets and continue their work on them. The name of the sheet to be opened has to be written in the textbox corresponding to the name.

F. **Interoperability -** Interoperability in SocialCalc with any file format means that files of that format can be read by the application. SocialCalcNet can currently open files with .wk4 and .xls(Excel sheets) extension. Users can upload these files from their system and then view them as SocialCalc sheet. The uploaded files are transferred to the server. The extension of the file uploaded, is retrieved from the file name. SocialCalc stores the sheet data in the form of a string. The data in the wk4 and XLS files is however not in the same format. Therefore to open those files in SocialCalc their contents is first converted to the SocialCalc string.

G. **Collaboration -** Collaboration is the feature wherein various users can work together on the same sheet, thus increasing work efficiency. The feature of collaboration in SocialCalcNet is still in the early development stages. The collaborated sheets are organized by defining the author of the sheet and the group name. These two are used together to identify a sheet uniquely. For secure collaboration, users are asked for a sheet-id that is specified by the sheet author at the time of creation.

SocialCalcNet uses JSON(JavaScript Object Notation) to make these RPC's [13]. While making a RPC all the arguments at the client side are packed into a string and then sent to the server. The server unpacks these arguments and calls the appropriate routine to handle these requests. The results are again packed in the form of a string and then send to the client.

There are mainly two functions that the application currently calls via RPC –

(i) Sending the changes to the server. Whenever a change in the current sheet is observed it is sent to the server for other users to collect. This change is sent in the form of a command string which is then used by other users to introduce these changes to their sheets as well. For this data transfer a record of all the active users working on a sheet is stored along with the changes sent by each of them.

(ii) Collecting the changes from the server and displaying them on the sheet. For this the application currently uses synchronous checking. The client synchronously keeps on checking for any data submitted at the server by making RPC at fixed intervals. If any new data is found, it is collected and then displayed in the sheet.

## V. ARCHITECTURE TO USE SPREADSHEET FRAMEWORK ON THE CLOUD

Once the basic collaboration over the cloud through the Spreadsheet was achieved, we devised architecture for using this Spreadsheet as a chassis for building several other Activities which require the support of a Spreadsheet. The diagram below shows this architecture in brief:

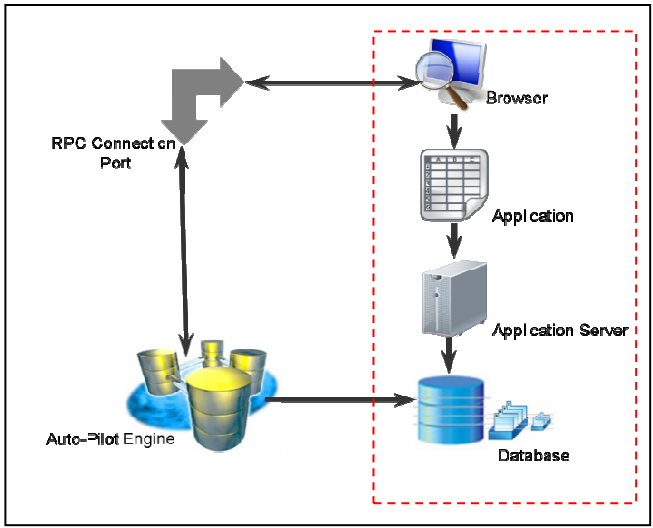

Fig 4: Proposed Architecture

In the architecture that has been proposed, the components which have been implemented in the previous sections have been marked in red. Now, in order to exploit this system for more robust and hard to implement use cases, we have introduced two new components:

(i) **RPC Connection Port:** RPC is a powerful technique for constructing distributed, client-server based applications. It is based on extending the notion of conventional or local procedure calling, so that the called procedure need not exist in the same address space as the calling procedure. The procedures may or may not be on the same system, but the RPC connection allows inter-process communication, without programmers actually coding details for the implementation. In our architecture, we introduced the RPC Connection in order to bring into picture parallelism that would be required by complex systems using this architecture. While each instance of the application would be requiring computation bandwidth and server storage space, it is not possible to do this on a single machine. In such a scenario, it is useful to implement RPC, such that the Server which acts as the RPC Connection port, handles all the computation, while the individual machines logging into the systems simply run the application remotely.

(ii) Also, in figure point systems, while running the data stream on the cloud, there is a threshold limit to the streaming rates. More often than not, this is far less than recommended speeds. In order to increase the speed of performance of a Cloud application, a local cache or mirror of the data and computation is maintained on this RPC Connection server. This comes into picture only when required, so that during the running of the application, only the important computation data is transferred instead of the entire data involved in the process.

(iii) **Auto-Pilot Engine:** Several data in the application requires pre-processing and analytics. Instead of relying on the local machine to do this computation, we use RPC Connection port, in tandem with an auto-pilot engine. This engine, performs group analysis of data, performs various analytics and operations on the data and predictions based on the grouping. This engine also performs the task of synchronizing the received data with the database and eventually, presenting it to the end-user in the right format.

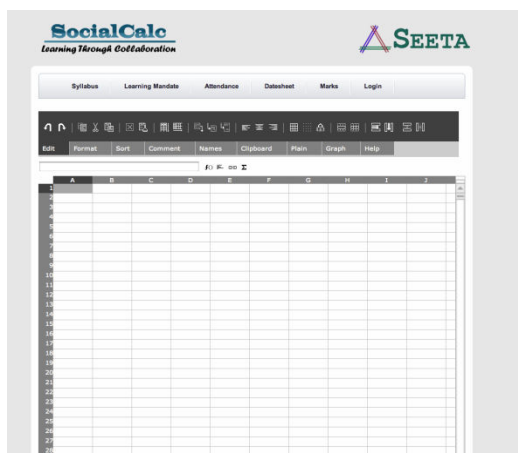

Figure 5: Spreadsheet on the Cloud

The most important use of these components is in their support to the existing Cloud Spreadsheet. While in the cloud version of the activity, several tasks such as collaboration and computation can be performed on the local machine itself, enhancing it in order to expand it to more complex systems would require support from back-end servers. This is where the RPC connection port and the auto-pilot engine contribute. These components ease the pressure from local machines by taking part in the computation process and transferring the same to the local machine through remote calls. The advantage of this lies in the increased efficiency and speed of the entire process on cloud.

## VI. WEB APPLICATIONS BASED ON SPREADSHEET

Once the Spreadsheet is running on the cloud, a combination of CSS and the coding of the Spreadsheet can be utilised in several scenarios to build complex web applications. The basic features such an exercise will utilise are:

- Computational prowess of Spreadsheet
- Design abilities of CSS
- Collaboration power of Spreadsheet
- Inter operability & cloud presence of Spreadsheet

This unique combination helps build **customized spreadsheet based websites**, which can be utilized to present complex data in simple formats, and at the same time, also be used in the context of collaboration.

One such use case which we have implemented is that of the School Administration system.

## VII. BUILDING THE SCHOOL ADMINISTRATION SYSTEM

The Administration system of an educational institution is a complicated system, which requires collaboration on various fronts. It requires students to be able to view their course work, as well as access Administration related modules such as Performance analysis, attendance management and work submission. Similarly, for the teachers, such a system should be an easy access to the Students' works, course material and administration modules such as Attendance management & performance analysis.

The School Administration system which we have built is based on these aspects. The basic framework is that of the Collaborative Spreadsheet SocialCalc.

The same framework is used to fetch the CSS outline for a design outlay spread over this Spreadsheet.

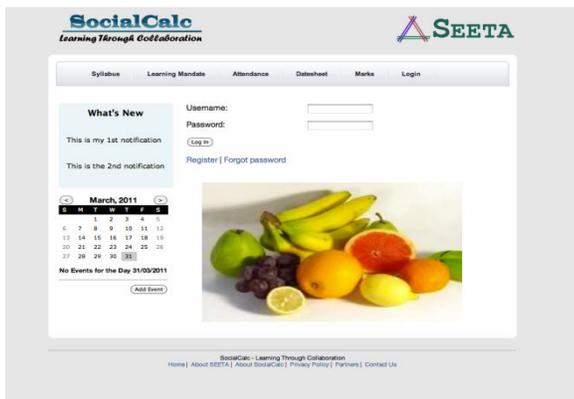

**Fig 6  Framework for Website**

Such a system enables us to make complete use of the Spreadsheet's exceptional abilities of computation & collaboration, at the same time, allows us to bring the design element associated with the website.

The various features included for students in the implementation include:

**Learning Mandate** – This section includes all lessons uploaded by the Teacher for the group. It also includes reminders on pending home-works and submissions. The mandate consists of theoretical chapters as well as specific tests and templates set by the teacher for these chapters.

Fig. 7   Framework for Teacher

Fig. 8   Framework for Student

1. **Attendance Management System** – This section includes the ability to fetch attendance for desired dates & submit leave applications.

2. **Calendar Management System** –This section is updated in real time with the latest happenings in school and the date-sheets regarding exams, results and events.

3. **Performance Management System** – This section consists of the history of records & grades obtained by the student. This is also refreshed in real time as and when the teacher uploads the Marks.

4. **Collaboration** – This feature is present throughout the website, wherein the student can simultaneously learn and collaborate with peers and teachers.

Similarly, such a system has also been designed for the Teachers and Administrators. This system also consists of:

1. **Learning Mandate**  - Where the teacher can upload the chapters & tests and assign them to group of students or classes. This section can also be utilised to set the Revision tests along with dates, which are then displayed as pending tasks in the students' landing pages. The Teacher uses this section to check the works submitted by students and mark them accordingly.

2. **Classroom Management System** – In this section, the teacher has the liberty to share her open spreadsheet with the class and utilise the collaboration tool to teach while working. She can also use this to observe while the students practise. She can also chat with any students who may be facing any problems during the class.

3. **Attendance Management System** - The Teachers can mark the attendances and submit the attendance to the Administrators for final evaluation.

4. **Calendar Management System** – This section is updated in real time with the latest happenings in school and the date-sheets regarding exams, results and events. The Teacher can utilise this section to send reminders to students about pending tasks, and this gets updated in the students' landing pages.

5. **Performance Management System** – The Teacher uses this section to evaluate the students and submit their Report Cards class wise. This is updated in real time in the Student's landing pages.

Hence, we can see that the basic Spreadsheet framework has been utilised to produce a web application for the complete school management. In a similar manner, several other applications can be built, which utilise the Spreadsheet on cloud to solve complex problems such as computation & fetching of data.

## VIII.   SCENARIOS FOR SPREADSHEET ON THE CLOUD

While Education and Enterprise are common use case scenarios for using a Spreadsheet activity on the cloud, it is more important to utilise the basic chassis for more complex situations. With the architecture defined above, composite applications, involving multiple users, huge databases and complex computations can be easily implemented. In the course of next section, we discuss two such useful cases, especially tailored for the Indian atmosphere:

(i) Health on Cloud – Implementing a Health record portal for Health professionals as well as patients, to interact and store data on the cloud.

(ii) Administration on Cloud – Implementing an Administration portal for Administrators as well as Consumers to interact and collaborate.

## IX. HEALTHCARE ON CLOUD

The power of on-the-go computing applied to healthcare systems has long since become a technology advancement to keep the eyes on. Frost & Sullivan predicts that with healthcarewill become a major trend in the coming years [14]. Healthcare providers are increasingly looking at automating processes at lower cost and higher gains; cloud computing can act as an ideal platform in the healthcare IT space. Google Health and Microsoft Health Vault are some examples of how the world is tuning itself to this new phenomenon.

With the architecture we discussed, we believe that the Spreadsheet as a framework can prove to be an extremely reliable background for building a good portal for Healthcare over the Cloud. In essence, this refers to making real time availability of patient and professional's data, seamless access of data between various hospitals and healthcare communities, ability for patients to upload new data and collaborate with online doctors for referrals, ability for professionals to consult other doctors through collaboration over the cloud by sharing patient data and a lot more. This not only reduces the cost of data storage for Hospitals, but also makes the interaction between professionals and patients seamless. Also, this helps in the standardization of patient data across the globe.

In the Indian context, with rural healthcare becoming a pressing issue with the day, it has become imperative to find solutions which will help organize this extremely unorganized sector. With the Cloud infrastructure in place and Information & Communication Technologies gradually creeping in to rural India, this could be the perfect solution bringing the rural population into the realms of technology. The issue of healthcare professionals not being able to devote time to this sector can also be reduced to a large extent, with collaboration and interactivity between professionals being made possible over the Cloud. Some of the features which we propose to be build into our Healthcare portal, based on the Spreadsheet framework are:

— Cloud interface for storing Healthcare data on the Cloud
— Separate interfaces for Doctors and Patients/Consumers
— A web interface for allowing Consumers to upload data
— A web interface for allowing Doctors to upload data
— Templates with pre-defined fields which can be directly loaded
— All data to be saved in Spreadsheet formats
— Ability to save profiles of patients in folders
— Ability to share files and folders with other Doctors only
— Ability to comment on a diagnosis which has been shared by another doctor
— Ability for Consumers to share their profiles/reports with Doctors they chose
— Collaboration to allow different Doctors to work on same patient simultaneously
— Collaboration to allow Consumers to view online Doctors closest to their location and allow them to chat and take advice

**USE CASE:** Consider a scenario in a rural village in India. With ICT in place, the local health clinic embarks on a drive and after a rigorous exercise organizes all the data related to all residents into Spreadsheet format, with the help of pre-defined templates. This data is stored on the Cloud, the only infrastructure deployed at the site being a basic computer. This data is now shared with all hospitals in the nearest vicinity. Whenever there is a situation when the low level healthcare professional based in the local clinic is unable to make a diagnosis, he simply sends a message to all these connected hospitals and shares the details of the patient. The doctors can now collaborate over the Cloud, chat, give individual diagnosis, which other doctors can comply or overrule and the same is now interpreted to the Local Doctor. In case of emergencies, the nearest Hospitals with available doctors can even send resources.

Consider another scenario, in a city, where ICT has penetrated well. The patient, being well aware of his conditions, has already uploaded the data on the portal. Now, requiring a consultation from a good doctor, he can first contact available doctors through the portal, share his data, await their initial diagnosis and if need be, arrange for an appointment.

In a further scenario, the mobile version of the application being GPS enabled can help in emergency situations, by raising an alarm about a patient and sharing his data immediately with the Hospital in the nearest vicinity and demanding resources.

Hence, with these use cases, we can see that a simple architecture, based a spreadsheet activity can be a step forward to solving the problem of inaccessible healthcare solutions.

## X.   ADMINISTRATION  ON CLOUD

In his Senate confirmation hearing in May 2009, AneeshChopra, USA's first chief technology officer (CTO), stated that cloud computingholds a number of advantages for the government. [15] These include "reduced cost,increased storage, higher levels of automation, increased flexibility, and higher levelsof employee mobility." More than anywhere else, India requires adopting these technologies with an immediate urgency in order to cleanse a system continually marred by allegations of corruption and inefficiency.

One such area we identified is the Public Distribution System or the PDS, which is a national program that distributes food grains at subsidized process to the poor in the country. The grains are procured from the Food Corporation of India (FCI) and distributed at the Fair Price Shops spread across the country.[16] Considering the numbers such as 14 million tonnes of grains, and 16 states amongst which these were distributed, it is a mammoth exercise which leads to immense operational problems. [17] For instance, out of the stated grains, only 5 million tonnes of grains actually reached the stipulated population. The rest was unaccounted for. While the Government is taking steps through Biometric systems and over hauling of the mechanism of distribution, it is an estimated huge spend on infrastructure.

With the architecture we defined in the precious sections, we propose a low cost system of administration on the Cloud. Some of the features are:

— Position SocialCalc as a "Collaborative Resource Management Tool"that ensures the proper distribution of grain and keeps real time track of:
   - the Amount of grains credited to each FCS
   - the amount of grain being distributed to consumers
   - the amount of grain that is unused

— Collaboration will allow hierarchal supervision of the entire process, reducing dependence on middlemen. Features include:

- Collaboration on same sheet to allow Real Time Tracking
- Chat Enabled Collaboration to allow interaction between Administration

— Multiple User entry points to allow all Administrators to Log in from different areas and access the Resource Management Tool

— Consumer Portal can allow users to enter their Ration Card number, view past transactions, get updates about next ration availability and amount of grains to be issued in their name

— In case a portal for consumers is not accessible, the consumer issues can be sorted by theFCS Officer by simply entering the Ration Card number and checking details.

— Administrator of local area can make information available to all Users regarding the Date of new Grain arrival, etc.

— Customers can also chat with Administrators to issue complaints or discuss grievances

— Issues and Problem identification is easier due to transparency. Any unnatural dealingscan be caught by the system and officers can be warned.

**USE CASE:** Consider a rural scenario in India. When the FCI releases the grain, the various Fair Price Shops (FPS) receive the grains [18]. The FPS manager then logs into the system and maintains a log of the amount of grain received and the number of customers he will be catering to. The district administration has access to records of all the FPS logs. Now, the district administrator can create schedules and upload them on the portal, informing both the FPS managers as well as customers about the dates during which the grains will be available. On the stipulated days, the FPS start distributing the grains and the Administrator keeps real time track of the logs.

A certain customer having missed the date, can either log onto the portal himself or through the FPS Officers, to check his previous transactions and intimates the manager about this. The manager also checks past transactions and immediately sanctions a new date for this customer to collect the grains he missed.

In this manner, the administration can be made transparent, accountable and customer friendly. The same system can be replicated for various other Governmental procedures.

### XI. FUTURE OF CUSTOMIZED WEB APPLICATIONS ON THE CLOUD

Technology in any form is ineffectual, unless transformed to suit the needs of the situation. Cloud computing, in various forms, can be customized to suit the Indian environment, and this paper was our small effort in achieving this. Through the paper, we discussed several scenarios, ranging from Education to Healthcare to Administration on the cloud. However, this is just the beginning. With a sound architecture and framework in place, the spreadsheet on cloud can be used a tool in several other scenarios. As we embark on this exciting journey, we hope to learn more and contribute more in development of web based applications.

# REFERENCES

[1] Gartner Inc. at Gartner Symposium/ITxpo, October, 2009. Available at http://www.gartner.com/it/page.jsp?id=1454221. Retrieved May, 2011.

[2] Available at: http://laptop.org/en/. Retrieved April, 2011.

[3] Available at: http://sugarlabs.org/, http://laptop.org/en/. Retrieved April, 2011.

[4] Dan is the co-creator of first spreadsheet application. More information could be found at: http://www.bricklin.com/. Retrieved April, 2011.

[5] Oeschger, I. & Turner, D., 2003. Creating XPCOM Components [Online], Available at: http://www.csie.ntu.edu.tw.

[6] Availableat: http://seeta.in/wiki/index.php?title=SocialCalcNet. Retrieved April, 2011.

[7] Available at: http://code.google.com/appengine/docs/whatisgoogleappengine.html. Retrieved April, 2011.

[8] Mark Chang , Jackson He , Enrique Castro-Leon, Service-Orientation in the Computing Infrastructure, Proceedings of the Second IEEE International Symposium on Service-Oriented System Engineering, p.27-33, October 25-26, 2006. Available at: http://portal.acm.org/citation.cfm?id=1174321. Retrieved April, 2011.

[9] Available at: http://en.wikipedia.org/wiki/Sandbox_%28computer_security%29. Retrieved April, 2011.

[10] Available at: http://www.djangoproject.com/. Retrieved April, 2011.

[11] Available at: http://code.google.com/appengine/docs/python/datastore/gqlreference.html. Retrieved April, 2011.

[12] Patrícia Gomes Soares, On remote procedure call, Proceedings of the 1992 conference of the Centre for Advanced Studies on Collaborative research, November 09-12, 1992, Toronto, Ontario, Canada

[13] Available at  http://www.json.org/. Retrieved April, 2011.

[14] Frost & Sullivan, December 2011. Available at http://www.slideshare.net/FrostandSullivan/mega-trends-that-will-shape-the-future-of-the-world . Retrieved April, 2011.

[15] The Wall Street Journal, Washington Wire. Available at http://goo.gl/pp1Pd. Retrieved April, 2011.

[16] Department of Food & Public Distribution. Available at www.fcamin.nic.in. Retrieved April, 2011.

[17] Goa Chamber of Commerce and Industry. Available at http://www.goachamber.org/cms/index.php?option=com_content&task=view&id=803&Itemid=20 . RetrievedApril, 2011.

[18] Available at: http://en.wikipedia.org/wiki/Public_distribution_shop. Retrieved April, 2011.